\documentclass[useAMS,usenatbib]{mn2e}
\usepackage{graphicx}

\title[The PMS star HD\,34282: A very short period $\delta$~Scuti-type
  pulsator]{The pre-main sequence star HD\,34282: A very short
  period $\delta$~Scuti-type pulsator}
\author[P.J. Amado et al.]{P. J. Amado$^{1}$\thanks{E-mail:
        pja@iaa.es}, A. Moya$^{1}$, J. C. Su\'arez$^{1,2}$,
        S. Mart\'{\i}n-Ruiz$^{1}$, R. Garrido$^{1}$,
        \newauthor
        E. Rodr\'{\i}guez$^{1}$, C. Catala$^{2}$ and
        M. J. Goupil$^{2}$\\
$^{1}$Instituto de Astrof\'{\i}sica de Andaluc\'{\i}a (CSIC), P.O. Box 3004,
E-18080, Granada, Spain\\
$^{2}$LESIA, Observatoire de Paris-Meudon, 92195 Meudon Principal
Cedex, France}

\date{Released 2004 Xxxxx XX}

\begin{document}

\date{Accepted. Received; in original form}


\maketitle


\begin{abstract}
HD\,34282 has been found to pulsate during a systematic search for
short-term photometric variability in Herbig Ae/Be stars with the goal
of determining the position and size of the pre-main sequence
instability strip.  Simultaneous Str\"omgren photometry is used in the
frequency analysis, yielding two frequencies with values of
$\nu_1=79.5$ and $\nu_2=71.3$ c\,d$^{-1}$.  The light curve with the
largest amplitude is that of the {\sl u} band.  This behaviour, which
is not common for $\delta$~Scuti stars, is explained as pulsation in a
high radial order in stars near the blue edge of the instability
strip. The main period, with a value of 18.12 min, represents the
shortest period observed so far for a $\delta$~Scuti-type pulsator.  A
seismic modelling, including instability predictions and rotation
effects, has been attempted.  It is found that both main sequence
and pre-main sequence models predict modes in the range of 56 to 82
c\,d$^{-1}$ (between 648 and 949 $\mu$Hz), corresponding to
oscillations of radial order $n$ from 6 to 8. The highest of the
observed frequencies only becomes unstable for models of low
metallicity, in agreement with results from spectroscopic
measurements.
\end{abstract}

\section{Introduction}

Pre-main sequence (PMS) stars with masses above 1.5 M$_{\sun}$ are
known as Herbig Ae/Be stars \citep{Her60, Str72b}.  These stars raise
several important questions which still need to be answered.  First,
their PMS nature needs to be confirmed, as their location in the
Hertzsprung-Russell (HR) diagram, clearly above the main sequence (MS)
\citep{Str72b, vdA98}, leads to the ambiguity that they could be
either PMS or post-MS objects.  Once their PMS nature is ascertained,
the Herbig Ae/Be stars can be used to constrain the modelling of PMS
evolution and to study their coupling with the circumstellar (CS)
environment, involving e.g. magnetic processes, accretion/ejection
processes, exchanges of angular momentum, etc...  The study of the
pulsations of these stars represents, therefore, a unique opportunity
to answer these questions. \citet{Sur01} have shown that some
non-radial unstable modes are extremely sensitive to the details of
internal structure, and make it possible to distinguish pre- from
post-MS stars, as well as to constrain the internal rotation.

PMS stars with masses above 1.5 M$_{\sun}$ are expected to cross the
instability strip on their way to the MS, spending typically 5\% to
10\% of their PMS phases within it \citep{Marc98}.  This time is
sufficiently long for a significant number of Herbig Ae stars to be
present in this strip, and, therefore, to presumably exhibit
$\delta$~Scuti-type pulsations.

Even though photometric variability induced by variable dust
obscuration or magnetic activity is rather high in these stars, the
time scales for these high amplitude variations are in principle
separated from those of $\delta$~Scuti-type pulsations: Keplerian
rotation of the CS envelope, presumably responsible for variable dust
obscuration, occurs on time scales of months, and the star's rotation,
at the origin of the variability due to surface activity, is typically
of the order of one to several days, while p-modes in these stars are
found with periods of minutes to hours.

The observations secured so far of the pulsational variability in
Herbig stars are not sufficient to verify the existence, the width and
the location of the instability strip in the HR diagram.  It is
therefore very important to start a systematic photometric survey of a
large number of Herbig stars, in order to study the observational
characteristics of the PMS instability strip, and to compare them with
theoretical predictions \citep{Marc98}.  This survey has already been
undertaken at the Sierra Nevada Observatory (OSN), Spain, within the
framework of a collaboration between the Observatoire de Paris and the
Instituto de Astrof\'{\i}sica de Andaluc\'{\i}a. Although still
ongoing, this survey has already yielded a first important result in
the discovery of four short-term variables, one of which is HD\,34282.

\begin{figure}
   \vspace{-4mm}
   \resizebox{\hsize}{!}{\includegraphics{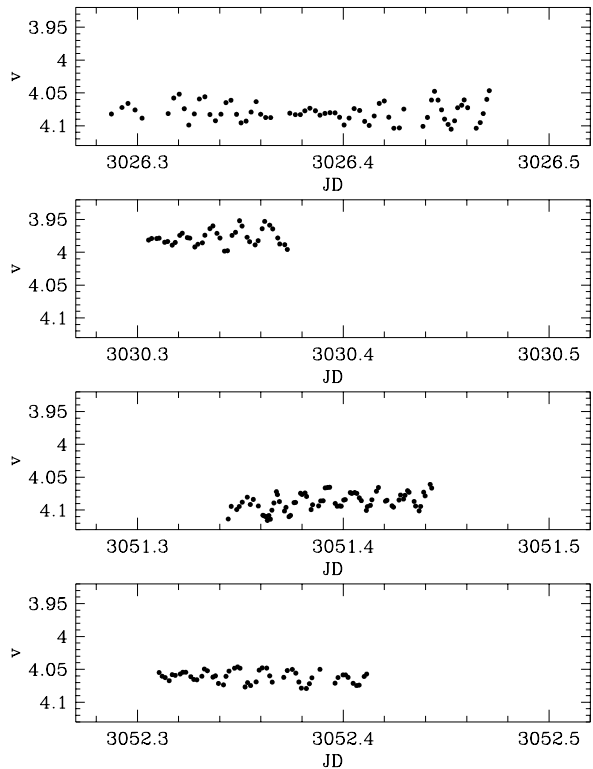}}
   \vspace{-9mm}
   \caption{{\sl v} light curves of the $\delta$~Sct PMS
            star HD\,34282 during four nights in January and February
   2004 (JD+2450000.00).}
   \label{Fig:vcurves}
\end{figure}

HD\,34282 (= V1366~Ori) has been classified in the spectral range
between A0 and A3 by several authors \citep{Gray98,Mora01,Meri04}.
This is a very interesting object because: 1) the star has strong IR
excess \citep{Sylv96,Mal98,Meri04}, 2) it is a relatively nearby star,
according to Hipparcos measurements $d=160^{+60}_{-40}$~pc,
\citep{vdA98} and 3) observations with the IRAM 30m antenna revealed a
double-peak profile in the $J=2\rightarrow 1$ transition of the CO
molecule, characteristic of a large rotating disk, further supported
by Plateau de Bure interferometer observations \citep{Pie03}.

\section{Data observations and reduction}

The star was observed during four nights in January-February 2004,
producing a time baseline of 25 days. The observations were collected
with the 0.9-m telescope at the OSN and the automatic six-channel
Str\"omgren spectrophotometer \citep{Niel83}.  Standard stars were
also observed to transform to the standard system.  The star was then
monitored for around 4 hours on the first night, and for two hours on
the remaining nights.  Figure~\ref{Fig:vcurves} shows the {\sl v}
light curves of HD\,34282 for the four nights of our observations.  The
ranges of both axes have been fixed in order to show the number of
hours of observations within one single day and the brightness
variations of the star from day to day.  The mean standard magnitudes
and indices for HD\,34282 are: $V=9.873$, $(b-y)=0.126$, $m_1=0.174$,
$c_1=1.001$ and $\beta=2.918$.

\section{Photometric variability}

\begin{figure}
    \resizebox{\hsize}{!}{\includegraphics{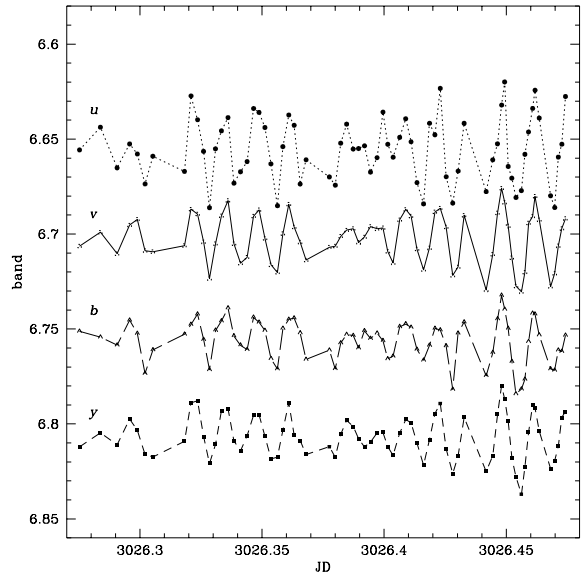}}
    \vspace{-9mm}
    \caption{{\sl u}, {\sl v}, {\sl b} and {\sl y} band light curves of
             HD\,34282 on the first night of observations. The mean
             values of the {\sl u} and {\sl y} light curves have been
             displaced for the sake of clarity.}
    \label{Fig:all_col}
 \end{figure}

Long-term variations were reported by \citet{Mal98} who measured an
optical variability of the order of 2.5 mag in the $V$ band, which
suggests that HD\,34282 might be a UX~Orionis star, i.e., a
pre-main-sequence star, typically of intermediate mass, which is
distinguished from other pre-main-sequence stars by its large
photometric and polarimetric variations, thought to be due to variable
extinction by CS dust.  Our data for HD\,34282 show photometric
variations at two time scales: from day to day and within a day, as
seen in Fig.~\ref{Fig:vcurves}.  The day-to-day variations are not the
main subject of the present work but rather the rapid intra-day
variations.  A blow-up of the data taken on the first night, where
these intra-day variations are clearly visible, is shown in
Fig.~\ref{Fig:all_col}.  The plot presents, from top to bottom, the
{\sl u}, {\sl v}, {\sl b} and {\sl y} light curves, showing that the
amplitude of the modulation decreases from the ultraviolet to the
yellow bands.

\begin{figure}
   \resizebox{\hsize}{!}{\includegraphics[angle=-90]{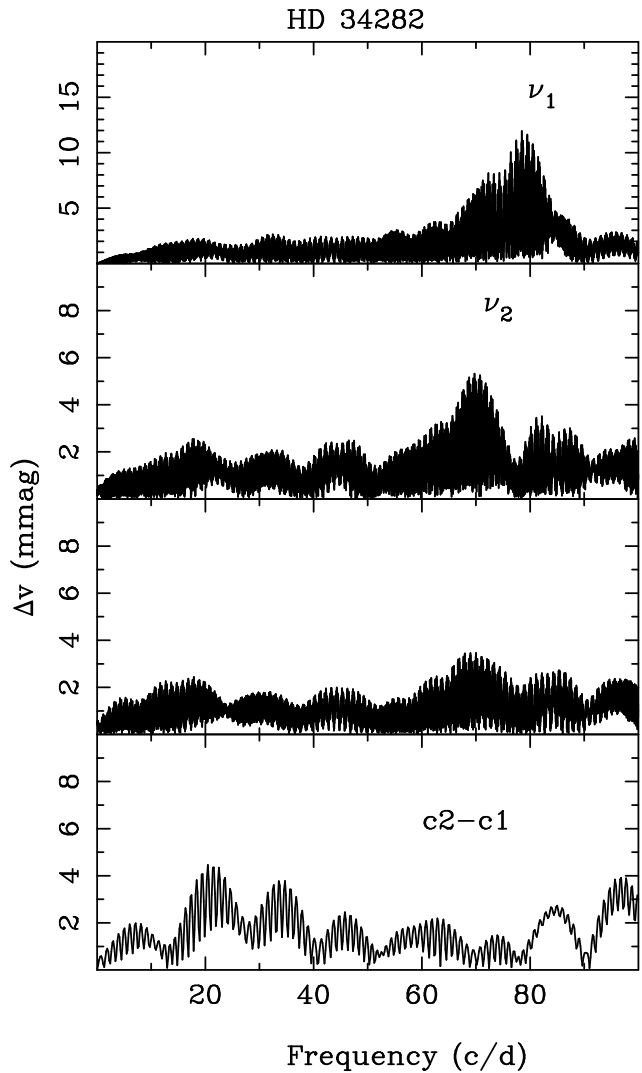}}
   \caption{Fourier analysis of the {\sl v} band Str\"omgren
            data. Results are given in Table~\ref{Tab:freq_list}.  The
            Fourier transform of the differential photometry of the
            comparison stars (C2$-$C1) is given in the bottom panel}
   \label{Fig:freq_anal}
\end{figure}

In order to analyse these rapid variations avoiding interference with
the day-to-day variations, the mean value for each day was removed.
This average subtraction is actually a low frequency filtering which
does not affect our analysis because the long-term variations only
show significant power in the low frequency region of the Fourier
spectrum up to 5 c\,d$^{-1}$, very far from the rapid variation
frequencies which fall in the region of 60-80 c\,d$^{-1}$.  The
frequency analysis performed for the {\sl v}-data, actually the best
S/N band, is shown in Fig.~\ref{Fig:freq_anal}.  The procedure is
similar to that presented in, e.g. \citet{Breg94}: pre-whitening of
simultaneous frequencies at each step until the S/N ratio decreases
below 4.  The last panel of Fig.~\ref{Fig:freq_anal} shows the Fourier
transform of the differential {\sl v} light curve of the comparison
stars.  The time span of these data is only of two days as compared to
the 25-d baseline of the variable's observations, hence the different
aspect of their Fourier spectra.

\begin{table}
\caption{Frequency in cycles per day, signal to noise, amplitude and
         its error and phase and its error for the two detected peaks
         in the Fourier spectrum of HD\,34282.}
\begin{tabular}{lcrrccc}
\hline
\multicolumn{2}{c}{Frequency}  &  \multicolumn{1}{c}{S/N} &
\multicolumn{1}{c}{Amp} & Error  & Phase & error\\
\multicolumn{2}{c}{c\,d$^{-1}$}  &  & \multicolumn{1}{c}{(mmag)} &
(mmag) & (rad) & (rad)\\
\hline
\multicolumn{2}{c}{{\sl v} band} &&&&\\
$\nu_1$ & 79.5 & 19.4 & 10.7 & 0.6 & 0.2 & 0.1\\
$\nu_2$ & 71.3 & 10.3 &  5.7 & 0.6 & 2.3 & 0.1\\

\multicolumn{2}{c}{{\sl b} band} &&&&\\
        & 79.5 & 20.3 &  8.2 & 0.6 & 0.4 & 0.1\\
        & 71.3 & 10.2 &  4.9 & 0.6 & 2.3 & 0.1\\

\multicolumn{2}{c}{{\sl y} band} &&&&\\
        & 79.5 & 14.6 &  8.2 & 0.6 & 0.4 & 0.1\\
        & 71.3 &  8.7 &  4.9 & 0.6 & 2.3 & 0.1\\
\hline
\end{tabular}
\label{Tab:freq_list}
\end{table}

The final frequency solution for each photometric band is presented in
Table~\ref{Tab:freq_list}.  Frequency uncertainties remain a matter of
controversy \citep[see, for instance,][]{Schw91} and the conservative
rule-of-thumb given by \citet{Loum78} was chosen: 1.5/T, where T is
our time baseline, yielding in our case 0.06 c\,d$^{-1}$.  The
oscillation periods of this object are the shortest so far found for a
$\delta$~Scuti variable.  Particularly the highest frequency
mode, with a period of only 18 min, can be considered as a new
world-record for this type of pulsating star. Up to now, the two
$\delta$~Sct variables with the shortest main periods known were the
pulsating primary components in the semi-detached Algol-type eclipsing
binaries RZ~Cas \citep[22 min,][]{Rodr04} and AS~Eri \citep[24
min,][]{Mkrt04}.  In both cases, the largest light curve amplitudes
are observed in the {\sl u} band as it seems to be the case for our
star.  This behaviour is explained as pulsation in a high radial order
in stars near the blue edge of the instability strip as discussed in
\citet{Rodr04}.

\section{Theoretical modelling}\label{sec:theormodel}

\begin{figure}
   \resizebox{\hsize}{!}{\includegraphics{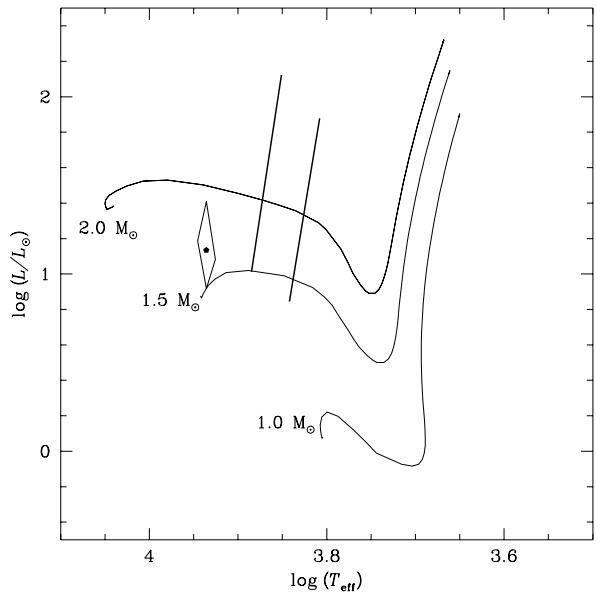}}
   \vspace{-7mm}
   \caption{HR Diagram for the PMS star HD\,34282.  The solid
            symbol and the error box indicate the position of the star
            given by the physical parameters of \citet{Meri04}.  The
            thick straight lines represent the PMS instability strip
            by \citet{Marc98}. The evolutionary tracks have been
            computed using the PMS models in the code {\sc cesam}.}
   \label{Fig:HRD}
\end{figure}

The code TempLogG \citep{Roge95} was used to determine the physical
parameters of the star from the standard Str\"omgren-Crawford indices
derived for this work.  The code classifies the star to be in the
spectral-type region A0-A3 and the resulting physical parameters are
$T_{\rm eff}=8760~{\rm K};~ \log{g}=4.4$.  This is correct if its
Str\"omgren indices are not `contaminated' with non-standard
photometric features coming from the surrounding material, such as
line emission and/or filling-in of the H$\beta$ line.  However, the
echelle spectra of \citet{Meri04} show a CS contribution in this
Balmer line among others, although not too large.  These same authors
give physical parameters for the star from high and medium resolution
spectra: $T_{\rm eff}=8625\pm200$~K, $\log{g}=4.2\pm0.2$ and ${\rm
[Fe/H]} = -0.8 \pm 0.1$.  These latter values were used to place the
star in the HR diagram (see Fig.~\ref{Fig:HRD}) and to produce
theoretical stellar models.  The solid symbol and the error box in
this figure give the position and errors for HD\,34282.  The
evolutionary tracks have been computed using the PMS models of the
{\sc cesam} evolutionary code \citep{Morel97}.  The straight thick
lines represent the PMS instability strip from \citet{Marc98} and, as
it can be seen, the star falls outside and towards the hotter side of
it.

Two different stellar models, one in the MS and another one at a PMS
stage, with physical parameters given by the position of the star in
the HR diagram, were computed using the {\sc cesam} code.  The
pulsational characteristics were calculated following the
prescriptions given in \citet{Moya04}.  The non-adiabatic oscillation
code provides not only predictions of the unstable modes but also
non-adiabatic quantities necessary to calculate phase differences and
amplitude ratios which can then be compared with those derived from
the observed Str\"omgren colour variations.  The MS model predicts
unstable radial and non-radial modes from 64.4 up to 81.4~c\,d$^{-1}$
(corresponding to radial orders from $n=6$ to 8), fitting fairly well
the observed frequency range (from 69 up to 80~c\,d$^{-1}$).  The PMS
model predicts unstable modes in a very similar range, from 56.3 up to
82.2~c\,d$^{-1}$, which also includes the observed periods.
Therefore, from the point of view of the excited modes, both models
are valid and no preference is deduced.

It must be said, however, that for the largest frequency to be
unstable, the low metallicity found for this star by \citet{Meri04} is
essential.  In the particular case of this star, a MS model with solar
metallicity has a mass of 1.9\,M$_{\sun}$, whereas a model with ${\rm
[Fe/H]} = -0.8$, as derived from the aforementioned spectroscopic
work, has a mass of 1.5\,M$_{\sun}$.  The main consequence of this
mass difference is the different range of unstable frequencies,
particularly for the higher radial orders.  The blue stability edge is
found to be around the same radial order $n$ in both cases, but for
the solar metallicity model the highest excited frequency is around
850~$\mu$Hz (73.4~c\,d$^{-1}$) whereas, for the spectroscopically
derived metallicity, it is slightly higher, around 950~$\mu$Hz
(82.1~c\,d$^{-1}$) for the MS and 1030~$\mu$Hz (89.0~c\,d$^{-1}$) for
the PMS one, being the observed frequency of 920~$\mu$Hz
(79.5~c\,d$^{-1}$).  For models with fixed $T_{\rm eff}$ and $\log{g}$
(the observational ones), the observed frequencies become excited only
for models of lower mass, which can only be achieved by decreasing the
metallicity.

\begin{figure}
   \resizebox{\hsize}{!}{\includegraphics[angle=-90]{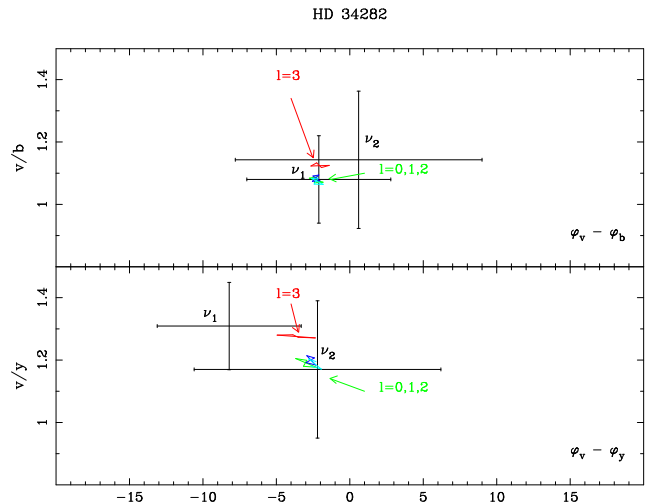}}
   \caption{Phase/amplitude diagrams for the Str\"omgren {\sl v} and
            {\sl b} (top) and {\sl v} and {\sl y} (bottom) bands
            showing the different loci for various $\ell$ numbers.
            Large crosses are observed values with their error bars.}
   \label{Fig:freq_id}
\end{figure}

The colour information provided by the Str\"omgren photometry enables
a further comparison with theoretical models through the use of the
phase differences and amplitude ratios between two photometric bands.
As shown in \citet{Garr90}, one of the best mode-discriminant indices
is that formed by the pair {\sl v\,--\,y}.  Using the non-adiabatic
quantities described in \citet{Moya04} a phase-amplitude diagram can
be constructed for the specific case of HD\,34282 and it is displayed
in Fig.~\ref{Fig:freq_id}.  Unfortunately, the discrimination is not
good for high order modes and high temperatures.  With the large error
bars derived for the present scarce observations, the oscillations are
compatible with all possible non-radial and radial modes up to
$\ell=3$.  Loci for different $\ell$-values were calculated taking
into account the uncertainties in effective temperature and surface
gravity and the dependence of the non-adiabatic quantities with the
oscillation period for both the MS and PMS models \citep[see][for a
detailed explanation]{Moya04}.

\section{Rotation}

From the seismological point of view, rotational velocities of the
order of, and even smaller than, that of HD\,34282 \citep[{\sl
v}\,$\sin{i}=110\pm10$~km\,s$^{-1}$,][]{Meri04} must be taken into
account in both the stellar structure modelling and the oscillation
computations for a correct interpretation of the observations
\citep{Saio81,DG92}.  Both aspects have been considered here and,
particularly, the adiabatic oscillation frequencies are corrected for
effects of rotation to second order, i.e. centrifugal and Coriolis
forces \citep[see][for a recent application]{Sua02aa}. Models are
computed for two rotational velocities (110 and 130~km\,s$^{-1}$) and
lying in the ranges $1.50$--$1.55$\,M$_{\sun}$ and $\log{T_{\rm
eff}}=[3.91,3.95]$, i.e., within the photometric errors.  From these
models, it can be concluded that rotation does not significantly
modify the range in which observed frequencies are found.
Furthermore, this range corresponds to radial order $n=[6,7]$, for
$\ell=0$, 1 and $2$ modes, thus lying within the range of predicted
unstable modes.  These results must be interpreted carefully since the
instability analysis does not take into account the effects of
rotation.

\section{Conclusion}

A systematic search for short-term photometric variability has been
undertaken for all known Herbig Ae stars with the goal of detecting
their intrinsic variability and of precisely locating them in the HR
diagram, in order to constrain the position and size of the observed
PMS instability strip.  The work presented here for HD\,34282 is one
result of this search.

HD\,34282 was found to pulsate at at least two frequencies.  This is
an important result in itself since it adds to the short list of PMS
stars known to belong to the $\delta$~Scuti pulsators.  Furthermore,
the highest of its frequencies yields a period of only 18 min, which
is the shortest found so far for this type of star. Moreover, the
largest amplitude of the light curves in the Str\"omgren system is
observed in the {\sl u} band.  This behaviour is explained as
pulsation in a high radial order in stars near the blue edge of the
instability strip.

The central star of HD\,34282 is theoretically expected to pulsate
whether it is a normal MS or a PMS star in the same position in the HR
diagram.  Furthermore, the highest of the observed frequencies only
becomes unstable for models of low metallicity, pointing thus towards
the same conclusion as that from spectroscopic measurements, that is,
that the star is metal-poor.

More observations better distributed in time are needed to allow the
detection of more periods and more precise phases and amplitudes, so
permitting a better understanding of this object.  This will give us a
possibility of modelling the physical conditions of the deep interior
of a PMS star.

\section*{Acknowledgments}

PJA acknowledges financial support at the Instituto de
Astrof\'{\i}sica de Andaluc\'{\i}a-CSIC by an I3P contract
(I3P-PC2001-1) funded by the European Social Fund. One of us wants to
thank the french PNPS (Programme National de Physique Stellaire) for
financial support. We also thank the anonymous referee for useful
comments and corrections which helped us to improve this manuscript.

\bibliographystyle{/home/dfe/pja/LATEX/STYLES/MNRAS/mn2e}
\bibliography{/home/dfe/pja/THESIS/mybib}

\bsp


\end{document}